\documentclass{IEEEtran}
\usepackage{amsmath,graphicx,subfigure}
\usepackage[usenames,dvipsnames,x11names]{xcolor}
\usepackage{tikz}
\usetikzlibrary{arrows,shapes}
\usepackage{enumerate}

\begin{document}
% 
% \title{}
% \author{}
% \maketitle

%\begin{abstract}

%\end{abstract}
\title{Communicating Under Channel Uncertainty}
\author{Naqueeb Warsi, Rahul~Vaze, Tapan Shah \\
Tata Institute of Fundamental Research\\
School of Technology and Computer Science\\
Homi Bhabha Road, Mumbai 400005\\
Email: \{naqueeb, vaze, tapan\}@tcs.tifr.res.in}
\date{}
\maketitle
%\markboth{To be submitted to IEEE Trans. on  Information Theory Draft Version 5}

\begin{abstract}
For a single transmit and receive antenna system, a new constellation design is proposed to combat errors in the phase estimate of the channel coefficient. The proposed constellation is a combination of PSK and PAM constellations, where PSK is used to  provide protection against phase errors, while PAM is used to increase the transmission rate using the knowledge of the magnitude of the channel coefficient. The performance of the proposed constellation is shown to be significantly better than the widely used QAM in terms of probability of error. The proposed strategy can also be extended to systems using multiple transmit and receive antennas. 
\end{abstract}

\section{Introduction}
Most practical wireless communication systems e.g. 3GPP-LTE \cite{3GPPLTE}, WIMAX \cite{Wimax} use quadrature amplitude modulation (QAM) assuming that the fading channel coefficient can be estimated up to an arbitrary level of precision. Acquiring accurate channel coefficient estimate \cite{Hassibi2003}, however, is challenging problem in practice, and requires sufficient resources, e.g. large training period and sophisticated signal processing at the receiver. The problem is even more complex at reasonable values of signal to noise ratios (SNRs), and more often than not there is a significant error in the channel coefficient estimate.

In this paper we consider a scenario where the estimator makes an error in estimating the phase of the channel coefficient, but estimates the magnitude of the channel coefficient  correctly. 
We assume no error in magnitude estimation, since the widely used QAM is robust to error in magnitude of the channel estimate, and no new constellation design is required for small errors in magnitude estimation. 
On the other hand, with error in the phase of the channel  estimate, we show that the maximum likelihood detector is an angle detector, and constellation points should be as far apart as possible to minimize the probability or error. Since the angle separation between different points of QAM is minimal,  using QAM in the presence of  error in the phase of the channel estimate can lead to large probability of error \cite{Ma}, (also shown using simulations in this paper). With error in the phase of the channel estimate, a natural strategy is use phase shift keying (PSK) that has the maximum angle separation. PSK, however, is most suited for the scenario when no information about the channel estimate is available (neither phase nor amplitude), %\cite{Hochwald1},
 and cannot take advantage of the knowledge of the magnitude of the channel coefficient.

\begin{figure}
\centering
\includegraphics[width=1.8in]{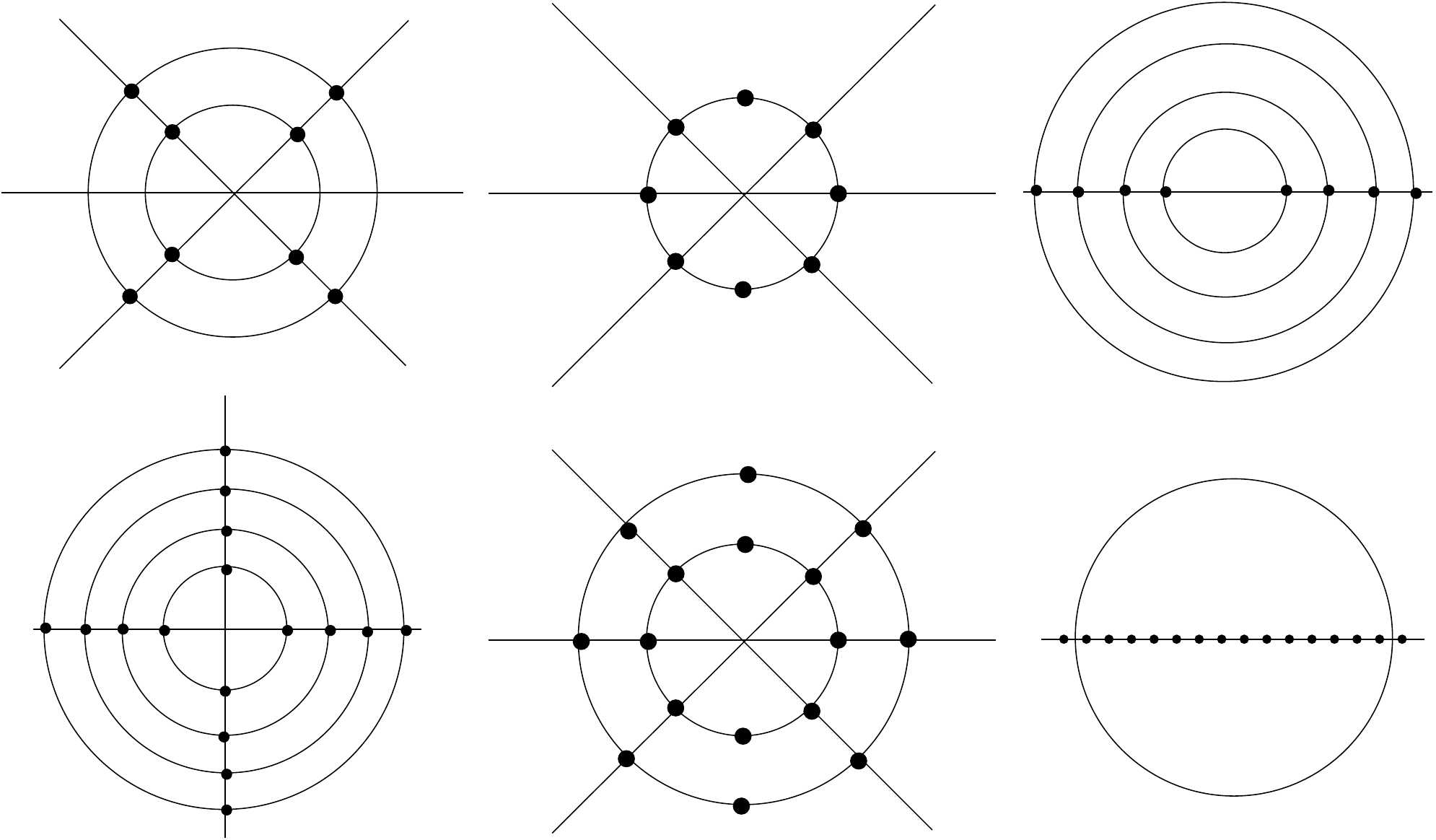}
\caption{8 points and 16 points constellation design.}
\label{fig:exm}
\end{figure}

To exploit the knowledge of the magnitude of the channel coefficient, we propose a signal constellation that is a combination of PSK and pulse amplitude modulation (PAM). For designing a constellation with $M$ points, the new constellation consists of $N$ concentric circles, with $K$ constellation points on each circle such that $KN=M$. In other words, there are $K$ lines with $N$ points on each line, where angle between each line is $\frac{2\pi}{K}$.
For example see Fig. (\ref{fig:exm}) where different configurations for $M=8$, and $16$, are illustrated. The idea behind this new constellation is that PSK is robust to error in the phase estimate of the channel coefficient, and one of the $K$ lines can be detected correctly even in the presence of error in the phase estimate of the channel coefficient. Assuming that the correct line has been decoded, using the exact knowledge of the magnitude of the channel estimate, decoding any point on the line is equivalent to PAM decoding. Thus, the decoding strategy is a two-step process, where in the first step, one of $K$ lines which minimizes the angle between the received signal is decoded, and then in the second step the point on the decoded line closest to the received signal in Euclidean distance is decoded as the transmitted signal. 
This combination of PSK and PAM performs better than QAM in terms of error probability in the presence of error in phase estimation, and  has better transmission rate than PSK. We note that the proposed concentric circles constellation is known in literature (concentric QAM or APSK) where distance decoding is used assuming that channel coefficient is known exactly \cite{BookProakis, concentric}. The novelty of this paper is in the proposed decoding algorithm that makes this constellation robust to phase estimation errors.

The probability of error for the proposed constellation design is a function of $K$ and $N$. 
Thus, for a given phase error, one needs to find the optimal value of $K$ and $N$ to minimize the probability of error, given the total number of points $M=KN$. For example it is easy to notice that with large phase error one needs to decrease $K$ thereby increasing $N$, while for small phase errors we can have large $K$ and consequently small $N$. We show that minimizing the probability of error is a min max problem over $K$ and $N$, and finding a closed form solution in general is hard. However, we show that using numerical integration and using brute force search (since the number of feasible points is not very large) we can find the optimal values of $K$ and $N$. We show that there is a significant  performance improvement by using our proposed constellation design over QAM even for small values of phase errors, and an arbitrarily large gain for large phase errors.

The proposed strategy to combat phase estimation errors can also be extended to the case of multiple transmit and receive  antennas as follows. With multiple transmit antennas, unitary space-time modulation is robust to channel uncertainty \cite{Hochwald2000b, Hochwald2000a, Zheng2002}. To exploit the knowledge of the channel magnitude, the columns of the unitary matrices are multiplied with constants $a_i, \ i=1,2,\ldots, L$, thereby increasing the rate of transmission. Once again a two-step decoding is employed. First, the unitary matrix  closest to the received signal in the chordal distance is decoded \cite{Hochwald2000a}, and then $a_i$ is decoded that lies closest to the received signal in the Euclidean distance.

An important application of the proposed constellation design is in the new paradigm of using low precision analog-to-digital/digital-to-analog  converters (ADC/DAC) for high speed/high bandwidth/high sampling rate transmission \cite{onkar}, \cite{Madhow}.
 With very high sampling rates only  $2-3$ bits are used for quantization, thereby drastically  degrading the quality of channel estimate. Therefore with low precision ADCs/DACs, employing our new constellation design can improve the error rate performance significantly, since it is tailor made for combating estimation errors.

\section{Notation}
\begin{enumerate}
\item $CN(\mu,\sigma^2$) $\rightarrow $ A complex normal distribution with mean $\mu$ and variance $\sigma^2$.
\item $N(\mu,\sigma^2)\rightarrow$ A normal distribution with mean $\mu$ and variance $\sigma^2$.
\item$C$ $\rightarrow$ Complex signal constellation.
\item $\Re$ $\rightarrow$ Represents the real part of a complex number.
\item $<a,b>$ $\rightarrow$ Dot product between two vectors $a$ and $b$.
\item$\dag$ $\rightarrow$ Complex conjugate.
\item $|h|$ $\rightarrow$ Amplitude of $h$.
\item$Q(x)$ $\rightarrow$ Complementary error function.
\item AWGN$\rightarrow$ Additive white Gaussian noise.
\item $Z^+$ $\rightarrow$ Positive integers
\item $E\{.\}$ $\rightarrow$ Expectation. 
\end{enumerate}

\section{System Model}
Consider a wireless communication link with a single transmit and receive antenna. The received signal $y$ is
\begin{equation}\label{sigmodel}
{y = \sqrt{P}hx+w},
\end{equation}
where $x$ is the transmitted signal with $E\{x^2\}\le 1$, $h$ is the channel coefficient, $w$ is $CN(0,1)$ distributed AWGN, and $P$ is the transmit power. To model the richly scattered fading channel we assume that $h$ is $CN(0,1)$ distributed. Let $h$ be represented as
\begin{equation}
h = |h|\exp^{j\theta},
\end{equation}
 where $\theta$ represents the phase of $h$. An estimate $\hat{h}$ of $h$ is used by the receiver to decode $x$ from $y$. In this paper we consider the case when the amplitude of $h$ is exactly known at the receiver, while the estimator makes an error $\phi$ in estimating the phase of $h$. We assume here that the phase estimation error $\phi$ is uniformly distributed over $[-a,a]$, and $\hat{h}$ is given by 
\begin{equation}
\hat{h} = {|h|}\exp^{j(\theta+\phi)}.
\end{equation}

\section{ML Decoding Rule with phase error in the channel estimation}
From (\ref{sigmodel}), given the channel estimate $\hat{h}$, the probability of receiving $y$ given that $x$ was transmitted is 
\begin{equation}
P(y|\hat{h}x)=\textit{E}_\phi(\frac{1}{\sqrt{2\pi}}\exp^{\{-(y-hx\exp^{j\phi})(y-hx\exp^{j\phi})^{\dag}\}}).
\end{equation}
With $z=yh^\dag x^\dag$,
\small
\begin{equation}
\begin{alignedat}{2}
P(y|\hat{h}x)&=\textit{E}_\phi(\frac{1}{\sqrt{2\pi}}\exp^{\{-(yy^\dag-zexp^{-j\phi} -z^\dag\exp^{j\phi}+|h|^2|x|^2)\}}),&& \\
 &=\frac{1}{2a}\int_{-a}^{a}{\frac{1}{\sqrt{2\pi}}}\exp^{\{-(yy^\dag-{\Re}\{z\exp^{-j(\phi)}\}+|h|^2|x|^2)\}}d\phi.&&
\end{alignedat} 
\end{equation}
\normalsize
\noindent Assuming that $|x|^{2}=\mbox{constant}$, $\forall$ $x$ $\in$ $C$(signal constellation), $P(y|\hat{h}x)$ is maximized if 
\begin{equation}
	\int_{-a}^{a}\exp^{\{\Re(yh^\dag x^\dag \exp^{-j\phi})\}}d\phi,
\end{equation}
is maximized over the interval $[-a,a]$. As integral is a limiting case of summation, we can write (5) as follows 
\begin{equation}
\lim_{k\rightarrow 0 }k\varSigma_{r = 0}^{n-1}\exp^{\{\Re(yh^\dag x^\dag \exp^{-j(-a+rk)})\}},
\end{equation}
where $n\rightarrow \infty$ as $k\rightarrow{0}$ and $nk$ is equal to $2a$.
Therefore, to maximize the above summation, we need to maximize $\Re(yh^\dag x^\dag \exp^{-j(-a+rk)})$, $\forall(-a+rk)\in[-a,a]$, over $C$, which is equal to maximizing $<y,\hat{h}x>$. Hence the ML decoding rule is

\begin{equation}\label{mlrule}
\max_{x\epsilon \textit{C}}<y,\hat{h}x>.
\end{equation}
Therefore to minimize the probability of error, the angular separation between the constellation points should be maximized. 
Note that  the angular separation between the constellation point in QAM is minimal, and PSK cannot exploit the knowledge of $|h|$. 
To combat phase errors and achieve higher transmission rate compared to PSK, in the next section we present a constellation design that is a combination of PSK and PAM constellations. The basic idea behind the constellation design is that  PSK is robust to phase errors in channel estimation, while PAM constellation can be used to increase the rate of transmission by exploiting the knowledge of $|h|$.

\section{Constellation Design}

\begin{figure}
\centering
\includegraphics[width=1.2in]{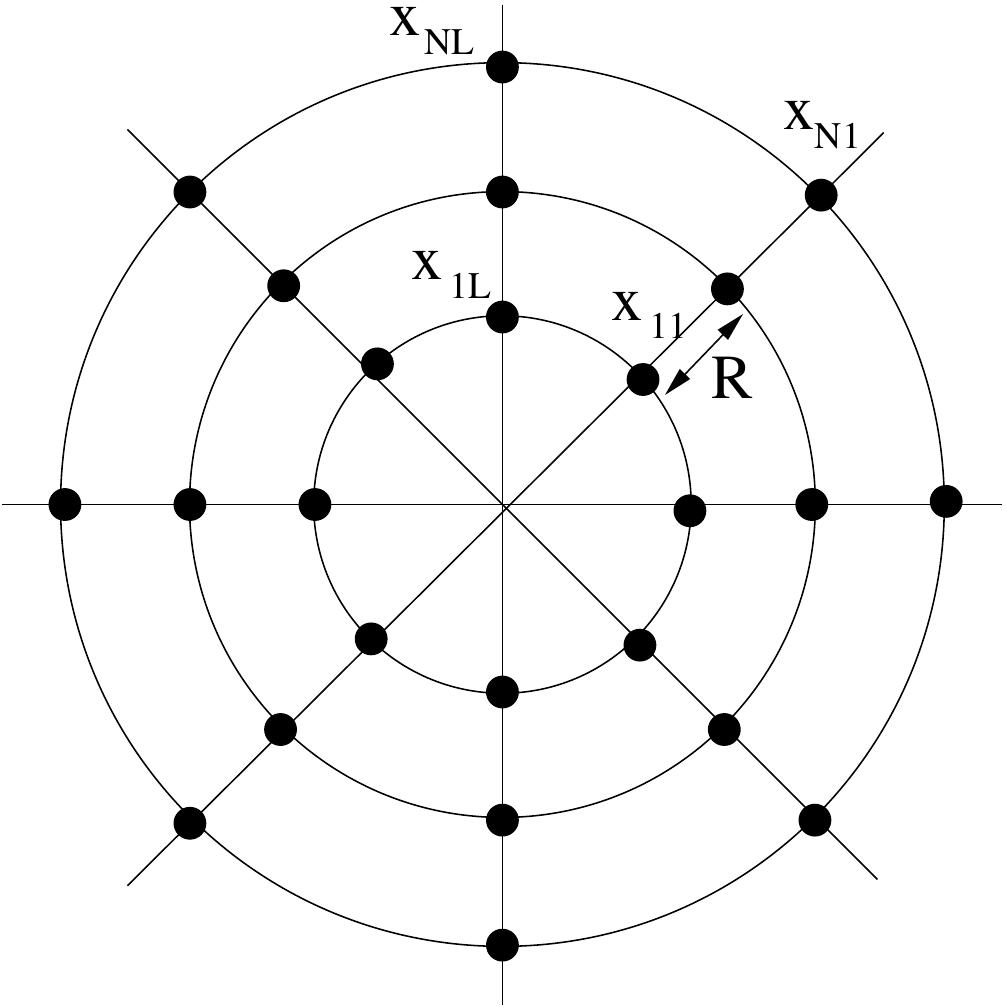}
\caption{A general PSK-PAM constellation design.}
\label{fig:genconst}
\end{figure}
\noindent Let $C$ be a complex signal constellation with signal set $\{x_1,\cdots,x_M\}$ with $M$ constellation points. In our proposed constellation design we divide the set $C$ into subsets $L_k$ where,
\begin{equation}
L_k = \{x_{ik}:angle(x_{ik})=\alpha_k\}, i = 1,\cdots, N, k = 1,\cdots,K,
\end{equation}
such that $M=KN$, where $|x_{ik}|=|x_{il}|$, $\forall k, l \in \{1,\cdots,K\}$,  $|x_{1k}|= R, \ \forall \ k$, and $R$ is also the Euclidean distance between two nearest elements of each subset, i.e. $|x_{(i-1)k}|=R+|x_{ik}|$, $\forall$ $k$. We define $\alpha_k$ as the angle of the subset $L_k$ made by $x_{ik}$, $\forall$ $i$, with the positive $X$ axis. The constellation design is illustrated in Fig. \ref{fig:genconst}. If the total number of subsets are $K$, and $\alpha_{k-1}<\alpha_k<\alpha_{k+1}$,$\forall$ $k$, then the angular difference between the subset $L_k$ and $L_{k+1}$ is given by $ \delta = \frac{2\pi}{K}$. With the power constraint of $E\{x^2\}\le 1$ and $M=KN$, $R$ can be written as,
\begin{equation}
 R = \sqrt{\frac{6}{M(\frac{M}{K}+1)(\frac{2M}{K}+1)}}.
\end{equation}
We represent our signal constellation as $(K,N)$ pair, where $K$ is the total number of subsets, and $N$ represents the cardinality of each subset. Fig. \ref{fig:exm} illustrates 8 points and $16$  points constellation designs for various values of $K$ and $N$ e.g. (4,2), (8,1), (2,4), 
(4,4), (8,2), etc.

\section{Decoding strategy}
The ML rule derived in the previous section is under the assumption that $|x|^2 =\mbox{constant}$, $\forall$ $x$ $\in$ $C$. However, in the proposed design this condition is not satisfied. To overcome the decoding problem, consider a constellation $C ^{*}$ where

\begin{equation*}
 C^{*} = \cup_{k=1}^{K} L_k^{*},
\end{equation*}

\noindent and $L_k^{*} = \{x_{1k}:x_{1k}\in L_k\}$, $k=1,\cdots,K$. Note that the constellation points of $C^*$ lie on a circle of radius $R$. The receiver now chooses an element $x_{1d}$ in $C^{*}$ which maximizes (\ref{mlrule}). Because the ML decoding rule is essentially an angle detector, if $x_{1d}$ $\in$ $C^{*}$ maximizes (\ref{mlrule}), it implies that $L_d$ is the most likely transmitted subset. After the above intermediate step, the receiver views the decoded subset as PAM constellation and decodes the received signal  $x_{nd}$ $\in$ $L_d$, which is closest to $y$ in Euclidean distance. The decoding strategy is therefore a two step process as summarized below.

\begin{enumerate}[Step 1$\rightarrow$]
 \item Decode the subset using the ML rule derived in the previous section. Let $L_d$ be the decoded subset where 
\begin {equation*}
 {d} = \mbox{arg max}_{k}<y,\hat{h}x_{1k}>,
k = 1,\cdots,K.
\end {equation*}
\end{enumerate}

\begin {enumerate}[Step 2$\rightarrow$]
\item After decoding the subset $L_d$ (Step 1) decode $x_{nd}$ $\in$ $L_d$ as the transmitted constellation point if 
\begin{equation*}
 n = \mbox{arg min}_i|y-\hat{h}x_{id}|,
i = 1,\cdots,N.
\end{equation*}
\end{enumerate}

\section{Error Analysis For the Proposed constellation design}
\noindent To derive the probability of error, let $x_{ik}$ $\in L_{k}$ be the transmitted signal.
The decoding error can occur because of the following two events,
\begin{enumerate}[Event 1$\rightarrow$]
 \item Subset $L_{d}\neq L_k$ is decoded by Step 1 of the decoding strategy.
 \item $L_{d}=L_{k}$, and point  $x_{jk}$ $\in$ $L_k$, $j\neq i$, is decoded by Step 2 of decoding strategy.
\end{enumerate}

\noindent The probability of error can be written as,
\begin{equation}
\begin{alignedat}{2}
 {P_e}^{total}=&\sum_{u=1,(u\neq k)}^{K}P\Bigl({L_d}\ = L_{u}|x_{ik}\in L_k\Bigr)+P\Bigl({L_d}=L_k|x_{ik}&&\\
&\in L_k\Bigr)
\sum_{j=1,(j\neq i)}^{N} P\Bigl(x_{nd}=x_{jk}|x_{ik},x_{jk}\in{L_k}\Bigr).&&
\end{alignedat}
\end{equation}

\noindent In the above equation $P({x_{nd}}=x_{jk}|x_{ik},x_{jk}\in {L_k})$ is the probability of the Event 2 i.e. $x_{jk}$ is decoded given that $x_{ik}$ is transmitted. Thus Event 2 occurs if
\begin{equation}
 |y-\hat{h}x_{ik}|>|y-\hat{h}x_{jk}|.
\end{equation}
If the maximum phase error $a$ satisfies
\begin{equation}
 \cos(a) \geq \frac{(2N-1)}{(2N)},
\end{equation}
then the probability of Event 2 is same as the probability of PAM decoding error given  by the formula \cite{Booktse}, and
\begin{equation}\label{pamdec}
 P\Bigl({x_{nd}}=x_{jk}|x_{ik},x_{jk}\in{L_k}\Bigr) = E_{h}Q\Bigl(\frac{|\hat{h}||i-j|R}{2\sqrt{0.5}}\Bigr),
\end{equation} where $|i-j|R$ represents the Euclidean distance between $x_{ik}$ and $x_{jk}$. We define $P\Bigl({x_{nd}}=x_{jk}|x_{ik},x_{jk}\in{L_k}\Bigr)$ as $P_{PAM}^{i\rightarrow j}$.
\\\textbf{Remark:} If the phase error does not satisfy (13), then re-estimate $h$ using the $y$ and the decoded $x_{1d}$ $\in L_d$ (step 1) i.e. $\hat{h} = \frac{y}{x_{1d}}$, re-estimating $h$ forces the phase error to satisfy (13).

Using (\ref{mlrule}), $P\Bigl({L_d}\ = L_{u}|x_{ik}\in L_k\Bigr)$ in (11) is the probability of the Event 1 i.e.
\begin{equation}
P \bigl ((<y,\hat{h}x_{1u}>-<y,\hat{h}x_{1k}>)\geq0|x_{ik}\in L_k \bigr ),
\end{equation}
where $x_{1k}\in L_k$ and $x_{1u}\in L_u$. 
Let $x_{ik} = iR\exp^{j\alpha}, x_{1k} = R\exp^{j\alpha}, \ i\in \{1,\ldots, N\}$, and  $x_{1u} = R\exp^{j\beta}$, then 

\begin{equation}
\label{dp1}
\begin{alignedat}{2}
 <y,\hat{h}x_{1k}>=&<|h|iR\sqrt{P}\exp^{\{j(\theta+\alpha)\}},|h|R\exp^{\{j(\theta+\phi+{\alpha})\}}>&&\\
 &+<w,|h|R\sqrt{P}\exp^{\{j(\theta+\phi+{\alpha})\}}>,&&
\end{alignedat}
\end{equation}
and
\begin{equation}
\label{dp2}
\begin{alignedat}{2}
 <y,\hat{h}x_{1u}>=&<|h|iR\sqrt{P}\exp^{\{j(\theta+\alpha)\}},|h|R\exp^{\{j(\theta+\phi+{\beta})\}}>&&\\
 &+<w,|h|R\exp^{\{j(\theta+\phi+{\beta})\}}>.&&
\end{alignedat}
\end{equation}

\noindent Subtracting (\ref{dp2}) from (\ref{dp1}) we get
%\small
\begin{equation}
\begin{alignedat}{2}
\bigl(<y,\hat{h}x_{1u}>-<y,\hat{h}x_{1k}>\bigr)=&|h|^2iR^2\sqrt{P}(a-b)+w_R|h|(c-d)&&\\
&+w_I|h|(e-f),&&
\end{alignedat}
\end{equation}
%\normalsize
\noindent where $w_R$ and $w_I$ represent the real and imaginary part of $w$, and $\{a,b,c,d,e,f\}$ are defined as follows
\begin{equation*}
\begin{alignedat}{1}
&a=\cos(\phi+\beta-\alpha),b=\cos(\phi),c=\cos(\theta+\phi+\beta),\\
&d=\cos(\theta+\phi+\alpha),e=\sin(\theta+\phi+\beta),f=\sin(\theta+\phi+\alpha).\\
\end{alignedat}
\end{equation*}
Let $ \lambda=w_R|h|(c-d)+w_I|h|(e-f)$. Note that $\lambda$ is Gaussian distributed with zero mean and variance $\frac{|h|^2((c-d)^2+ (e-f)^2)}{2}$, since $w_R$ and $w_I$ are independent and $N(0,1/2)$ distributed.
Therefore\begin{equation}
\begin{alignedat}{2}
P\Bigl({L_d}\ = L_{u}|x_{ik}\in L_k\Bigr)=\textit{E}&_{|h|\phi}\Bigl(\lambda > |h|iR^2\sqrt{P}(a-b)\Bigr).
\end{alignedat}
\end{equation}
With $N(\mu,\sigma^2)$ distributed  $\lambda$, 
\begin{equation*}
 P(\lambda \geq x)=Q\Bigl(\frac{x-\mu_\lambda}{\sigma_\lambda}\Bigr).
\end{equation*}
Hence after some manipulations
\begin{equation}
\label{linedec}
 P\Bigl({L_d}\ = L_{u}|x_{ik}\in L_k\Bigr)= E_{|h|\phi} Q\Bigl(\sqrt{2}|h|iR^2\sqrt{P}\sin(\phi+\frac{\beta-\alpha}{2})\Bigr).
\end{equation}

\noindent We define $P\Bigl({L_d}=L_{u}|x_{ik}\in L_k\Bigr)$ as $P_{Subset}^{k \rightarrow u}$. Then ${P_e}^{total}$ can be represented as 

\begin{equation}
 {P_e}^{total} \leq \sum_{u=1,(u\neq k)}^{K}P_{Subset}^{k\rightarrow u}+\sum_{j=1,(j\neq i)}^{N} P_{PAM}^{i\rightarrow j}.
\end{equation}

Note that $\sum_{u=1,(u\neq k)}^{K}P_{Subset}^{k\rightarrow u}$ and $\sum_{j=1,(j\neq i)}^{N}$ $P_{PAM}^{i\rightarrow j}$ in (23) are primarily governed by  $P_{Subset}^{k\rightarrow(k+1)}$ and $P_{PAM}^{i\rightarrow(i+1)}$  i.e. the probability of wrong decoding is governed by the nearest neighbor of $x_{ik}$ both in angular distance and Euclidean distance. 
Using (\ref{pamdec}) and (\ref{linedec}), $P_{PAM}^{i\rightarrow(i+1)}$ and $P_{Subset}^{k\rightarrow(k+1)}$ can be written as follows

\begin{equation}
 P_{PAM}^{i\rightarrow(i+1)} = E_{h}Q\left(\frac{|h|R}{2\sqrt{0.5}}\right),
\end{equation}
and
\begin{equation}
 P_{Subset}^{k\rightarrow(k+1)}\approx E_{|h|\phi} Q\left(\sqrt{2}|h|\sqrt{P}\sin\left(\phi+\frac{\delta}{2}\right)\right).
\end{equation}

Therefore, to minimize the probability of error we need to find 
\begin{equation}\label{designrule}
\begin{alignedat}{2}
&\min_{K,\frac{M}{K}} \max(P_{Subset}^{k\rightarrow (k+1)},P_{PAM}^{i\rightarrow (i+1)})&&\\
& \mbox{such that }\{M, \ \frac{M}{K} \in Z^+\} \mbox{ and } M=KN.&&
\end{alignedat}
\end{equation}
Hence the constellation design problem (finding $N$ and $K$) requires solving (\ref{designrule}). The difficulty in solving  (\ref{designrule}) is that it is a non-linear optimization problem, and moreover calculating $P_{Subset}^{k\rightarrow (k+1)}$ in closed form is difficult. However, $P_{Subset}^{k\rightarrow (k+1)}$ can be easily computed using numerical integration, and we take this approach for plotting the analytical results and use brute force search to find the optimal $N$ and $K$.

\section{Design Principles}
To minimize the probability of error we need to find $N$ and $K$ that minimizes (\ref{designrule}) for a fixed $M$. Finding $N$ and $K$ for a general case is hard and requires numerical integration and brute force search. Next we discuss some special cases for which $N$ and $K$ can be found easily. 
\begin{itemize}
\item {\bf Very small error in phase estimation, small $a$:}
In this case 
$P_{PAM}^{i\rightarrow j} \approx P_{Subset}^{k\rightarrow u}$. 
Hence to minimize $P_{e}^{total}$, the constellation design should have $N\approx K$.

\item {\bf Large  error in phase estimation, large $a$:}
In this case, to minimize $P_{e}^{total}$  we need to maximize the angular separation $\delta$ between two consecutive subsets. Therefore the constellation design should have $N> K$.

\item {\bf Unknown $h$:} When there is no information about the channel coefficient, i.e. neither the phase nor the amplitude of $h$ is known, then PAM decoding is not possible and thus $N$ has to be  $1$. Therefore we get the special case of PSK constellation where $K=M$.
\end{itemize}

\section{Simulation Results}
\begin{figure}[h!]
\vspace{-0.3cm}
\begin{center}
\includegraphics[scale=0.35]{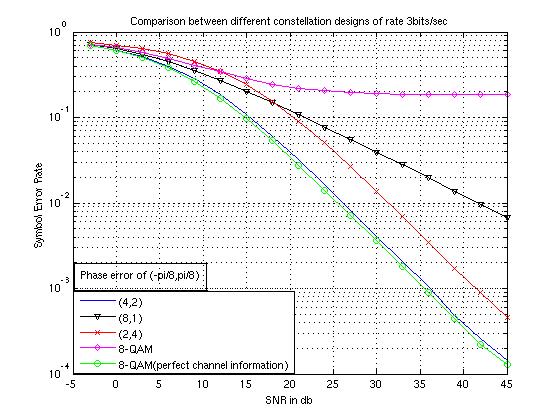}
\vspace{-0.3cm}
\caption{Performance results for different $3$ bits/sec constellation designs for a phase error of $(-\frac{\pi}{8},\frac{\pi}{8})$.}
\label{8QAM}
\end{center}
\vspace{-0.6cm}
\end{figure}

\begin{figure}[h!]
\begin{center}
\includegraphics[scale=0.35]{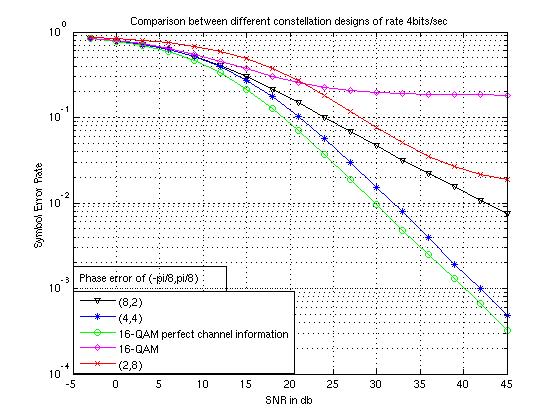}
\vspace{-0.3cm}
\caption{Performance results for different $4$ bits/sec constellation designs for a phase error of $(-\frac{\pi}{8},\frac{\pi}{8})$.}
\label{16QAM}
\end{center}
\vspace{-0.6cm}
\end{figure}

\begin{figure}[h!]
\begin{center}
\includegraphics[scale=0.35]{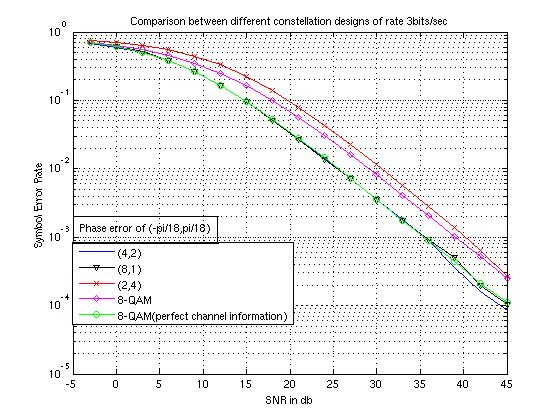}
\vspace{-0.3cm}
\caption{Performance results for different $3$ bits/sec constellation designs for a phase error of $(-\frac{\pi}{18},\frac{\pi}{18})$.}
\label{8QAM1}
\end{center}
\vspace{-0.6cm}
\end{figure}

\begin{figure}[h!]
\begin{center}
\includegraphics[scale=0.35]{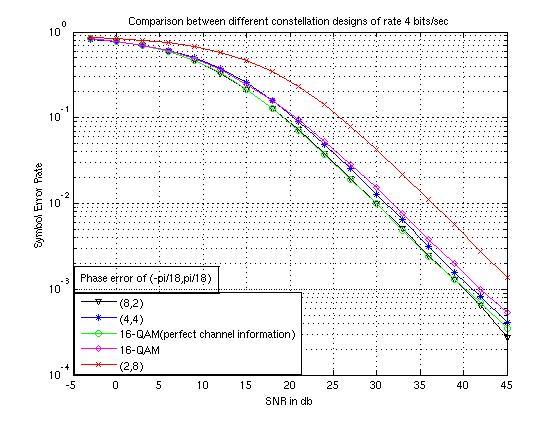}
\vspace{-0.3cm}
\caption{Performance results for different $4$ bits/sec constellation designs for a phase error of $(-\frac{\pi}{18},\frac{\pi}{18})$.}
\label{16QAM1}
\end{center}
\vspace{-0.6cm}
\end{figure}

\begin{figure}
\centering
    \subfigure[]{
        \includegraphics[scale=0.375]{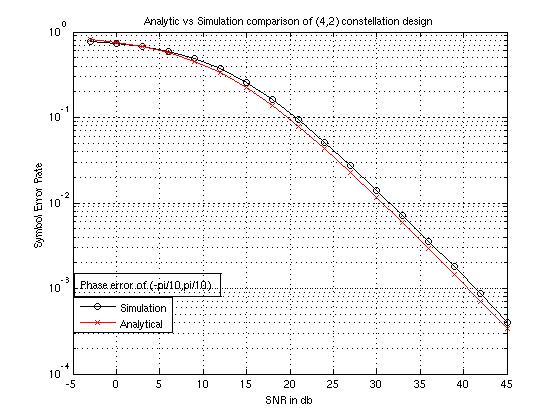}
        \label{a1}}
     \subfigure[]{
	\includegraphics[scale=0.375]{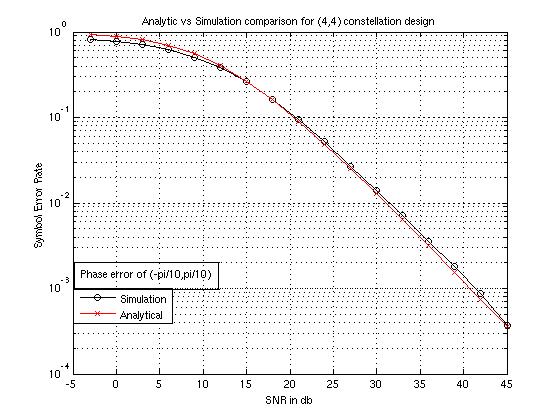}
         \label{b1}}
\caption{\subref{a1} Analytic vs Simulation comparison for $(4,2)$ constellation designs for a phase error of $(-\frac{\pi}{10},\frac{\pi}{10})$, \subref{b1} Analytic vs Simulation comparison for $(4,4)$ constellation designs for a phase error of $(-\frac{\pi}{10},\frac{\pi}{10})$.}
\label{Analytic}
\vspace{-0.3cm}
\end{figure}

We now present simulation results to demonstrate the performance improvement of the proposed constellation design over $8$-QAM and $16$-QAM with phase estimate error. Figs. (\ref{8QAM}) and  (\ref{16QAM}) compare the symbol error rate (SER) for  different values of $N$ and $K$ for $NK=8$ and $16$ i.e. for rate $3$ bits/sec and $4$ bits/sec with $8$-QAM and $16$-QAM as a function of SNR for phase error range of $(-\frac{\pi}{8},\frac{\pi}{8})$ i.e. $a=  \frac{\pi}{8}$. 
Simulation results show that for $a=  \frac{\pi}{8}$ (large phase error) the best constellation to use is $(4,2)$ for $M=8$, and $(8,2)$ for $M=16$, since the angular separation with $K=4$, and $K=8$, is $\frac{\pi}{2}$, and $\frac{\pi}{4}$, respectively,  which can easily tolerate a phase error of $\frac{\pi}{8}$.  Our results show that the new constellation design achieves a significant SNR gain over $8$-QAM and $16$-QAM constellation. We also consider relatively small phase error range of  $(-\frac{\pi}{18},\frac{\pi}{18})$ in  Figs. (\ref{8QAM1}) and  (\ref{16QAM1}), to show the considerable performance improvement of our strategy compared to $8$-QAM and $16$-QAM constellations. We also compare the probability of error generated using numerical integration and Monte Carlo simulation in 
Fig. (\ref{Analytic}), where we see that the difference between the two curves is minimal.

\section{Discussion}
Under the assumption that there is error in estimating the phase of the channel coefficient, we  derived the ML decoding rule and showed that the widely used QAM is not robust to error in  phase estimation of channel coefficient. Then we proposed a hybrid PAM-PSK constellation, which is more tolerant to errors in phase estimate as compared to QAM, and has a transmission rate higher than PSK. We calculated the error probability for the proposed hybrid PAM-PSK design and provided some design principles. Simulation results show that our approach results in a significant SNR gain (for large phase error)  as compared to the QAM constellations.

\bibliographystyle{IEEEtran}
\bibliography{IEEEabrv,Research}

\end{document}